\documentstyle[12pt]{article}
\title{A New Regularization Method in 3-Dimensional Momentum Space}

\author{Liang-gang Liu$^{1,2}$, Xiang-qian Luo$^{1,2}$, Wei Chen$^{2}$ \\
        $^{1}$CCAST (World Laboratory) P. O. Box 8730, Beijing, 100080 \\
        $^{2}$
        Department of Physics, Zhongshan University \\
        Guangzhou 510275, P. R. China }
\begin{document}
\maketitle
\vspace{2cm}
\begin{abstract}
We propose a new method to calculate the 4-dimensional divergent integrals.
By calculating the one loop integral as an example, the regularization 
of the integrals in 3-dimension momentum space are given in details. We find
that the new method gives the same results as the traditional dimensional
regularization method gives, but the new method has the advantage that it 
gives the real and the imaginary part separately.

PACS number(s): 11.10.Gh
\end{abstract}
\newpage
\section {Introduction}
It is well known that the dimensional regularization method is a powerful
and elegant tool to calculate and regularize the divergent integrals in 
4-dimensional energy-momentum space$(^{1,2})$. This method is widely used
both in particle physics as well as in nuclear physics$(^{3,4})$ in 
calculating the loops integration and renormalization. But in some cases,
such as of the zero-point fluctuation energy 
 $\int_\infty d{\bf p} \sqrt{{\bf p}^2 + m^2}$, where $m$ is 
the mass of nucleon or mesons, can not be calculated can not be calculated
by using the dimensional regularization formulae directly. In this case, a
cutoff momentum is introduced to truncate the upper limit of the integration
to make the divergence controllable$(^5)$. In our studies, we found this kind 
of 3-dimensional divergent integrals can be regularized by a new method
$(^6)$. In this paper, we will demonstrate that this method is also 
applicable to calculate other divergent integrals in 4-dimension 
energy-momentum space.

In the next section, we will show briefly the traditional dimensional regularization
of the integrals $F_1$ and $F_2$. In sect. 3, we give the details of the 
calculation of $F_1$, $F_2$ by our new method. A summary is given in the 
last section.

\section{4-Dimensional regularization of $F_1$ and $F_2$}
In the one loop level of vacuum fluctuation, the polarization insertion to 
the vertex function or propagators can be expressed in term of two functions
$(^7)$:
\begin{eqnarray}
F_1 = {i\over (2\pi)^4} \int_\infty d^4q {1\over q^2 - m^2 + i \epsilon},
\end{eqnarray}
\begin{eqnarray}
F_2(k^2, m^2_1, m^2_2) = {i\over (2\pi)^4} \int_\infty d^4q 
{1\over (q^2 - m^2_1 + i \epsilon)[(q - k)^2 - m^2_2 + i\epsilon]},
\end{eqnarray}
here $k, q$ are four energy-momentum.\footnote {We follow the convention of
J. D. Bjorken {\scriptsize AND} S. D. Drell, {\it Relativistic Quantum
Fields}, McGraw-Hill, New York, (1965).} The subscript $\infty$ indicates
the integration is in the whole 4-dimensional space. By use of the dimensional
regularization technique$(^{1, 2, 3})$, $F_1$ and $F_2$ are readily 
calculated, the results are follows:
\begin{eqnarray}
F_1 = - {m^2\over 16\pi^2} (\Gamma (\epsilon) - ln {\hskip 0.1cm}m^2 + 1 
+ \bigcirc (\epsilon)),
\end{eqnarray}
\begin{eqnarray}
F_2(k^2, m^2_1, m^2_2) = - {1\over 16\pi^2} (\Gamma (\epsilon) - 
\int^1_0 dx {\hskip 0.1cm} ln M^2_k(x) + \bigcirc(\epsilon)),
\end{eqnarray}
\begin{eqnarray}
 M^2_k(x) = k^2 x^2 - (k^2 + m^2_1 - m^2_2)x + m^2_1 ,
\end{eqnarray}
where $\epsilon \rightarrow 0^+, \Gamma (\epsilon)$ is gamma function,
$\Gamma (\epsilon) \sim {1\over \epsilon}$ so $F_1, F_2$ are divergent.
Notice that $M^2_k(x)$ is not positive-definite, provided that $k^2 > (m_1
+m_2)^2$, which is the threshold that the imcoming particle can decay into
particles $m_1, m_2$. Thus a imaginary part of $F_2$ appears,
\begin{eqnarray}
{\mbox Im}F_2(k^2, m^2_1, m^2_2) = - {1\over 16\pi} 
{\sqrt{\Delta} \over k^2} {\hskip 0.1cm} \theta(k^2 > (m_1 + m_2)^2),
\end{eqnarray}
\begin{eqnarray}
\Delta = [k^2 - (m_1 + m_2)^2] [k^2 - (m_1 - m_2)^2].
\end{eqnarray}

\section{Regularization in 3-dimensional space}
We propose in the following a new regularization method in 3-dimensional
space. Taking $F_1$ and $F_2$ as examples to demonstrate that the results are
the same as above.

\subsection{Regularization of $F_1$}
The denominate in eq. (1) can be expressed as $(q_0 - E_{\bf q} + i\epsilon)
(q_0 + E_{\bf q} - i\epsilon)$, here $E_{\bf q} = \sqrt{{\bf q}^2 + m^2}$. By 
use of Cauchy theorem, the integration of $q_0$ in the complex $q_0$ plane
gives:
\begin{eqnarray}
F_1 = {1 \over 2(2\pi)^3} \int_\infty {d{\bf q} \over E_{\bf q}}.
\end{eqnarray}

Now we assume the 3-dimensional integration is performed in n-dimension with 
the limit $n \rightarrow 3$, so $d{\bf q} \rightarrow d^n{\bf q} =
 q^{n-1} d{\hskip 0.1cm}q d\Omega_n$ with $\int_\infty d\Omega_n = 4\pi$
in the limit $n \rightarrow 3$, here $q \equiv |{\bf q}|$. Then eq. (8) 
can be reduced as follows:
\begin{eqnarray}
F_1 & = & {1 \over 8 \pi^2} (m^2)^{{n-3}\over 2} 
\int^\infty_0 d{\hskip 0.1cm}t 
{\hskip 0.1cm}t^{{n\over 2} -1}{\hskip 0.1cm}(1+t)^{-{1\over2}} \\ \nonumber
&  & = {1 \over 8 \pi^2} (m^2)^{{n-3}\over 2} \frac{\Gamma ({n\over2})
{\hskip 0.1cm} \Gamma({1\over2} - {n\over 2})}{\Gamma ({1\over2})}.
\end{eqnarray}
Expand it in a Laurent series about $\epsilon = {3-n\over 2}$, it gives a
result which is the same as eq. (3).

\subsection{Regularization of $F_2(k^2, m^2_1, m^2_2)$}
$F_2(k^2, m^2_1, m^2_2)$ can be rewritten as follows:
\begin{eqnarray}
F_2(k^2, m^2_1, m^2_2) = {i\over (2\pi)^4} \int^1_0 dx {\hskip 0.1cm} 
\int_\infty d^n{\bf q}{\hskip 0.1cm} dq_0 {1\over[q^2_0 - {\bf q}^2 
 - M^2_k(x)]^2}.
\end{eqnarray}
Since $M^2_k(x)$ can be both positive and negative, we should locate 
the poles in the denominator of eq. (11). The analysis shows that
\begin{eqnarray}
& M^2_k(x)& >0 {\hskip 0.3cm}{\mbox for}{\hskip 0.3cm} k^2 < (m_1+m_2)^2, 
\\ \nonumber
&{\mbox or}&, {\hskip 0.3cm}
k^2 > (m_1+m_2)^2 {\mbox and}  {\hskip 0.4cm} 0<x<x_1, \\ \nonumber
&{\mbox or}&, {\hskip 0.3cm}
k^2 > (m_1+m_2)^2 {\hskip 0.3cm}{\mbox and}{\hskip 0.3cm}x_2<x<1, 
\\ \nonumber
& M^2_k(x)& <0 {\hskip 0.3cm}{\mbox for} {\hskip 0.3cm} 
k^2 > (m_1+m_2)^2 {\hskip 0.3cm}{\mbox and}{\hskip 0.3cm} x_1<x<x_2,
\end{eqnarray}
where
\begin{eqnarray}
x_{1,2} = {1\over 2k^2}(k^2 + m_1^2 - m^2_2 \mp \sqrt{\Delta}).
\end{eqnarray}

Then eq. (11) can be written as follows:
\begin{eqnarray}
F_2(k^2, m^2_1, m^2_2) = {i\over (2\pi)^4}\{ \theta[k^2<(m_1+m_2)^2]
\int^1_0 dx {\hskip 0.1cm} \int_\infty d^n{\bf q}{\hskip 0.1cm} dq_0 
{1\over(q^2_0 - E_{\bf q}(x)^2 + i\epsilon)^2}
\\ \nonumber
+ \theta[k^2>(m_1+m_2)^2](\int^{x_1}_0dx + \int^1_{x_2}dx) \int_\infty
d^n{\bf q}{\hskip 0.1cm} dq_0 {1\over(q^2_0 - E_{\bf q}(x)^2 + i\epsilon)^2}
\\ \nonumber
+ \theta[k^2>(m_1+m_2)^2] \int^{x_2}_{x_1}dx 
\int_\infty d^n{\bf q}{\hskip 0.1cm} dq_0
{1\over(q^2_0 - {\bf q}^2(x) + |M^2_k(x)|)^2}\},
\end{eqnarray}
here $E_{\bf q}(x)^2 = {\bf q}^2 + M^2_k(x)$. It can be verified that 
\begin{eqnarray}
\int_\infty d^n{\bf q}{\hskip 0.1cm} dq_0
{1\over(q^2_0 - E_{\bf q}(x)^2 + i\epsilon)^2}
& = &\int_\infty d^n{\bf q}{\hskip 0.1cm} dq_0
{1\over(q^2_0 - {\bf q}^2(x) + |M^2_k(x)|)^2} \\ \nonumber
& = & i\pi^2 (\Gamma (\epsilon) - ln {\hskip 0.1cm}M^2_k(x) 
+ \bigcirc (\epsilon)).
\end{eqnarray}
Substitute eq. (14) into eq. (13) we obtain
\begin{eqnarray}
F_2(k^2, m^2_1, m^2_2) & = &
- {1\over 16\pi^2} (\Gamma (\epsilon) - 
\int^1_0 dx {\hskip 0.1cm} ln |M^2_k(x)| + \bigcirc(\epsilon))
\\ \nonumber
& + & {i\over (2\pi)^4}\cdot (-i\pi^2)\cdot (x_2-x_1)\cdot i\pi
\theta[k^2>(m_1+m_2)^2] \\ \nonumber
& = & - {1\over 16\pi^2} (\Gamma (\epsilon) - 
\int^1_0 dx {\hskip 0.1cm} ln |M^2_k(x)|) \\ \nonumber
& - &{i\over 16\pi}{\sqrt{\Delta} \over k^2} {\hskip 0.1cm} 
\theta[k^2 > (m_1 + m_2)^2] + \bigcirc(\epsilon).
\end{eqnarray}
We can see that it is the same as eq. (4) and eq. (6), therefore, demonstrate
that 3-dimensional regularization gives the same results as traditional
4-dimensional regularization method. This method can be extended to calculate
other divergent integrals, such as zero-point fluctuation energy mentioned in 
the Introduction. The calculation gives
\begin{eqnarray}
-{1\over (2\pi)^3} \int_\infty d{\bf q} \sqrt{{\bf q}^2 + m^2} &\rightarrow &
-{1\over (2\pi)^3}\int_\infty d^n{\bf q} \sqrt{{\bf q}^2 + m^2} 
\\ \nonumber
& = & -{1\over 32\pi^2} m^4 (\Gamma(\epsilon) - ln{\hskip 0.1cm}m^2 +{3\over2})
+ \bigcirc (\epsilon),
\end{eqnarray}
which is again the same as that given in ref. $(^8)$.
\section{Summary}
We have demonstrate by calculating integrals $F_1$ and $F_2$ that the new                                                                                                     
3-dimensional regularization method gives the same results as those the
traditional 4-dimensional regularization method gives. We also show that the 
new method can be used to calculate the zero-point fluctuation energy of
relativistic particles. In fact, this method can be generalized further to 
evaluate the quantum effect in two or one dimensional systems, it has a wide
application perspective in physics.
\vspace{2cm}
\section*{References}
\begin{description}
\item
$(^1)$ G. 'tHooft {\scriptsize AND} M. Veltman, Nucl. Phys., {\bf B44}
       189 (1972).
\item
$(^2)$ P. Ramond, {\it Field Theory: A Modern Primer}, Addison-Wesley,
       Redwood City, (1981).
\item
$(^3)$ B. D. Serot, J. D. Walecka, {\it Recent Progress in Quantum 
       Hadrodynamics}, to be published in the International Journal of 
       Modern Physics {\bf E}.
\item
$(^4)$ L. G. Liu, W. Bentz {\scriptsize AND} A. Arima, Ann. Phys.,
       {\bf 194}, 387 (1989).
\item
$(^5)$ M. Nakano, et al, Phys. Rev., {\bf C55}, 1 (1997).
\item
$(^6)$ L. G. Liu, {\it Quantum Effects in Nuclear Matter}, Ph. D. thesis,
       University of Tokyo, (1989).
\item
$(^7)$ J. A. Mignaco, E. Remiddi, IL Nuovo Cimento, {\bf 1A}, 376 (1971).
\item
$(^8)$ S. A. Chin, Ann. Phys., {\bf 108}, 301 (1977).
\end{description}

\end{document}